# Beam Halo Imaging with a Digital Optical Mask*

H.D. Zhang, R.B. Fiorito†, A.G. Shkvarunets, R.A. Kishek

Institute for Research in Electronics and Applied Physics
University of Maryland, College Park, MD

C.P. Welsch

University of Liverpool and Cockcroft Institute, Daresbury, UK

Abstract

Beam halo is an important factor in any high intensity accelerator. It can cause difficulties in the control of the beam, emittance growth, particle loss and even damage to the accelerator. It is therefore essential to understand the mechanisms of halo formation and its dynamics in order to control and minimize its effects. Experimental measurement of the halo distribution is an important tool for such studies. In this paper, we present a new adaptive masking method that we have developed to image beam halo, which uses a digital micro-mirror-array device (DMD). This method has been thoroughly investigated in the laboratory using laser and white light sources, and with real beams produced by the University of Maryland Electron Ring (UMER). A high dynamic range (DR~$10^5$) has been demonstrated with this new method and recent studies indicate that this number can be exceeded for more intense beams by at least an order of magnitude. The method is flexible, easy to setup and can be used at any accelerator or light source. We present the results of our measurements of the performance of the method and images of beam halos produced under various experimental conditions.

*Work supported by ONR and the DOD Joint Technology Office
†Corresponding author: rfiorito@umd.edu

# 1. Introduction

Beam halo is typically observed in particle beams [1, 2]. There is no well-accepted, rigorous definition of halo, but it is usually described as a low intensity distribution of particles observed at large radii from a more intense centralized portion of the beam, i.e. the "core". Beam halo is associated with emittance growth and thus its presence signifies decreased beam quality. More seriously, halo particles traveling far from the center of the beam can be lost or produce other undesirable effects [3, 4]. For a high energy particle beam, the lost particles can cause nuclear activation, damage to beam line components and an increase in radiation background. For positively charged beams, the secondary electron emission from the impact of halo particles with the beam pipe can result in an electron cloud around the beam, which complicates the beam dynamics [5, 6]. Moreover, since halos are present in the beam phase space, simple apertures that can remove halo particles at a particular beam line location are not effective [7].

A number of theories have been developed and simulation studies have been performed [8 -10] to describe the formation of beam halo. These studies have shown that there are many factors which can cause beam halo [11], e.g. intra-beam scattering, collective instabilities, misalignments, magnet errors, noise and resonances associated with both intrinsic incoherent processes and collective space charge forces. However, despite a wealth of theoretical and simulation studies, the mechanisms controlling the formation of halo and their interactions are not well understood. Moreover, few experimental studies of halos have been performed to guide and enlighten theory and simulation.

Experimental studies of halos are themselves hampered by the high dynamic range, DR $>10^5$, typically required to meaningfully characterize the halo particle distribution. Several approaches have been used to measure halo. These include 1) mechanical devices such as a wire scanners and beam scrappers, which are used to detect halo particles electronically or detect them via induced radiation [3, 12]; 2) indirect or secondary emission monitors, which are used to measure particles or induced radiation produced when the beam interacts with residual or purposely injected gas [13] in the accelerator; and 3) direct imaging of beam produced optical radiation that is linearly proportional to the beam charge distribution, e.g. phosphor light, transition, synchrotron, edge and undulator radiation. In this paper we focus our attention on the direct imaging of the halo distribution.

Imaging techniques are usually limited by the dynamic range of the camera sensor (e.g. CCD cameras) to several orders of magnitude. To extend the DR, a number of strategies have been suggested. One of these [14] is to initially use a neutral density (ND) filter to reduce the core intensity so that it does not saturate the camera sensor. Then the filter is removed, allowing the core intensity to saturate the camera sensor, in order to view the less intense wings of the distribution. However, this method is problematic because the light of the bright core is scattered in the optics as well as in the sensor, which can contaminate the measurement of the true halo distribution. There is an additional problem with sensitive sensors such as CCD's and especially intensified CCD's. Blooming in the CCD (charge spill to neighboring pixels) and electron avalanche in the intensifier can disrupt, or at least influence, the measurement and possibly damage the sensor [15]. Advanced charge injection device (CID) cameras do not suffer blooming and have the advantage of large intrinsic dynamic range (DR > ~$10^5$) [16]. However, these devices require very long acquisition times to achieve a very high dynamic range, cannot deal with very high light intensities and, in any case, light scattering and diffraction of light in the input optics can limit and compromise the measurement of the halo.

The systematic reduction of diffraction has been successfully addressed in another imaging approach, i.e. coronagraphy applied to beam halo imaging [17]. This method usually uses transmissive optics and a fixed size a blocking mask to filter out the central area of the beam image, highly polished lenses to avoid scattering and special apertures (Lyot stops), which partially filter out the diffraction effects produced by the input lens and the blocking mask itself. After these steps are applied, the halo is made more visible by increasing the exposure time of the camera. A DR > $10^6$ has been achieved with this method.

We have taken the idea of beam core masking a step further step by employing a commercially available digital micro-mirror device to generate a reflective mask. The primary advantages of this technique are: 1) the blocking mask produced on the DMD array can be simply programmed to conform (adapt) to an arbitrary shape of the beam core; 2) the intense light from the core can be rejected with very high efficiency, i.e. reflected at a large angle away from the optical path before reaching the camera optics, which avoids light scattering inherent to the standard coronagraph technique; and 3) the method can, in principle, achieve a high effective dynamic range with a modest dynamic range, low cost CCD camera. The effect of diffraction and other optical effects such as scattering and aberrations, which contribute to the point spread function (PSF) of the imaging system are still present and can, in principle, limit the measureable dynamic range. However, as we will see below, the real (measured) PSF of the optical system does not limit the DR of our measurements.

DMD technology has been highly developed by Texas Instruments Inc. (TI) and is used primarily in HD displays and projectors. DMD's are also available for research and development purposes in kit form. These are available with a number of different DMD formats and mirror reflectivities. Each kit includes a DMD, controller board, USB interface and software for computer control of the device. In the experiments described in this paper we use the DMD 'Discovery 1100' and 'Discovery 4100' kits manufactured by Texas Instruments Inc. [18]. Both of these include a XGA format, 14.4 mm × 10.8 mm DMD consisting of 1024 × 768 pixels. In the 1100, the DMD is integrated together with the controller board; in the 4100 the DMD is connected to the controller via a ribbon cable. Fig. 1.1(a) shows a picture of the 1100 device and Fig. 1.1 (b) shows an enlarged 3D drawing of one of the 13.68 μm × 13.68 μm aluminized silicon micro mirrors in the array [19]. Each pixel can be individually addressed electronically by a CMOS substructure and rotated ±12° about the diagonal to an 'on' or 'off' state, when a positive or negative voltage is applied to electrodes located beneath the two opposite corners of the micro mirror. In the 'on' state, incident light is reflected in a direction 24º with respect to the incident rays, while in the 'off' state, the light is directed 48° away from this path. Thus, by instructing specific micro-mirrors to 'flip' the DMD can be used as a programmable spatial filter.

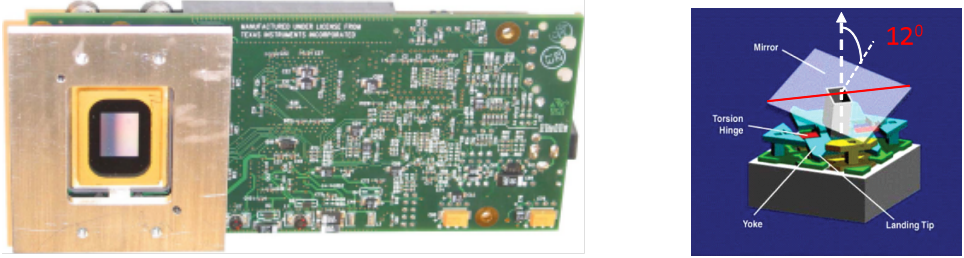

Figure 1.1 (a): Discovery 1100 DMD device, (b): Mechanical drawing of individual micro-mirror [18].

The filtering ability of DMD and its capability to produce a high effective dynamic range has been demonstrated by using it to measure the intensity profile of a low power He-Ne laser with a standard 8-bit CCD camera [19]. The experimental setup to do this is illustrated in Fig. 1.2. Note that without filtering (i.e. all pixels on) the spatial distribution of the core of the laser can only be measured within the inherent dynamic range (DR~100) of the CCD sensor. However, by using the DMD to generate a core blocking mask that is defined by establishing an intensity threshold, the light from the laser beam core can be reflected away from the optical path to the camera so that the rest of the profile, i.e. the wings or 'halo' of the distribution can be measured. By successively lowering the threshold to define the mask and adding the images (integration) the full dynamic range of the sensor can be used for each masked image.

Using this technique, a series of overlapping 2D intensity images of the laser beam distribution were obtained. Line scans across each of these images is shown in Fig. 1.3. To view all the line scans on one logarithmic plot, the intensities are scaled by the exposure time of each masked image and the unmasked plot is normalized by the peak intensity of the laser. Note that the intensity range of each line scan is limited to approximately two decades, which corresponds to the dynamic range of the 8 bit CCD camera used to make the measurements. The variable $\sigma$ in Fig. 1.3 represents the value of the variance, assuming that the laser profile is a Gaussian distribution. The resultant composite profile has an effective DR ~ $10^5$.

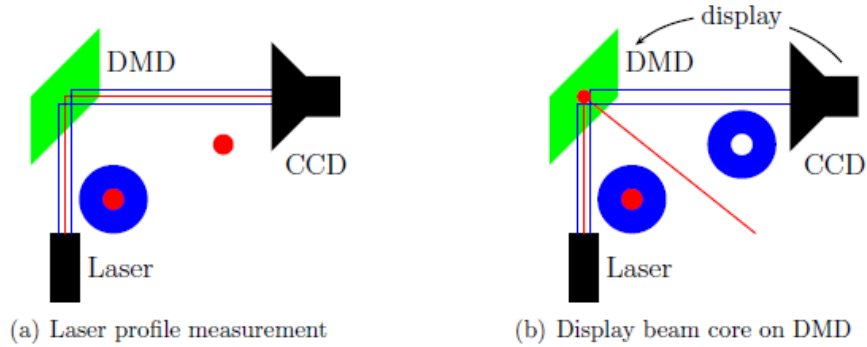

Figure 1.2: Laser profile measurement using DMD [19].

These laser studies indicated that the DMD masking method could potentially be more useful than a fixed spatial mask, such as used in the coronagraphic technique described above, and could serve to image the halo of a charged particle beam with a high dynamic range.

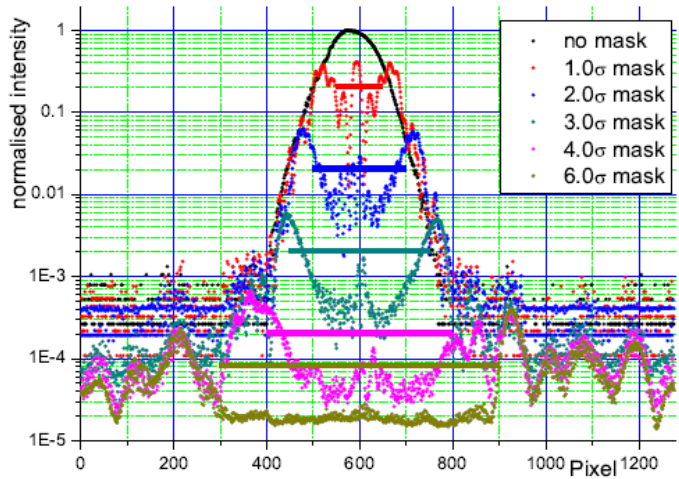

Figure 1.3: Normalized scans of laser images with various mask sizes; color horizontal bars indicate the mask size [19].

In the course of these measurements, it was also observed that the DMD produces a cross like diffraction pattern, which is characteristic of an array of rectangular apertures, when illuminated by the laser; i.e. the DMD acts like a 2D grating. And the question was raised about how this might effect the image and measurement of beam halo. However, our initial experiments, which used the DMD in a real imaging system with non-monochromatic light, showed that the diffraction effects were not significant [20].

In this paper, we report the results of more detailed measurements that verify our previous results. We discuss these and other properties of the DMD, which require optical compensation in order to effectively use the device in an imaging system. We describe the details of our optical system, the algorithm we have developed to spatially filter the beam image with the DMD, our measurements of the dynamic range and performance of the imaging system and present the results of our latest measurements of beam halos at the UMER accelerator.

We emphasize that the halos we have observed were empirically generated [21] and cannot be presently understood in terms of theoretical models. Therefore, no in depth analysis of the physics of the halos is attempted here. This will only be possible when systematic studies of halo generation are undertaken. The results presented in this work are intended only to illustrate the technique and demonstrate the optical performance of the halo imaging system we have developed. Our UMER results and preliminary results recently obtained at the JLAB 100 MeV Energy Recovery Linac [22] indicate that a DR $\sim 10^5$ can be achieved with this technique and that it may be possible to increase this DR by more than a order of magnitude.

## 2. Experimental Technique

2.1 Generic optical imaging system

Based on the properties of DMD described in above, we have designed a flexible imaging system that can be readily applied to image beam halo at any accelerator. Here the 'light source' is any radiation that can be used to image the beam. The essential features of the design are schematically shown in Fig. 2.1.

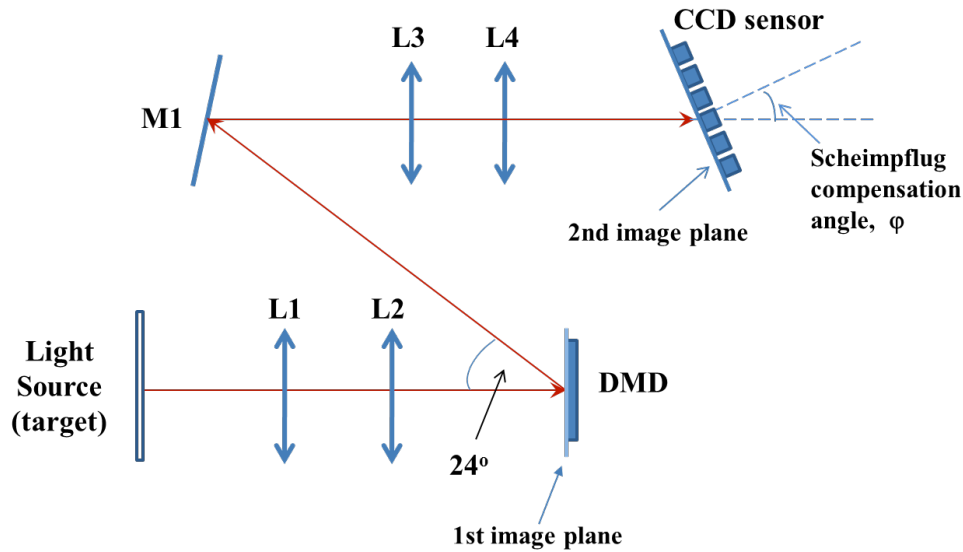

Figure 2.1: Schematic of halo imaging optics using DMD.

The setup shown in Fig. 2.1 can be considered as two optical channels: 1) the source, lenses L1, L2 and the DMD surface, which is perpendicular to the optical axis and is the first image plane; and 2) the DMD, considered as a new source, lenses L3, L4 and the CCD sensor, which is the second image plane. The lenses pairs L1, L2 and L3, L4 are used to adjust the magnification and focusing in channels 1 and 2, respectively. Two rotational compensations are required: 1) the DMD must be rotated by $45^0$ about the optical axis; 2) the camera sensor must be rotated in the horizontal plane by the Scheimpflug angle. The details of these compensations will be described later in this chapter.

2.2 Characteristics of UMER

In order to test the DMD imaging method described above with a real beam, we use the University of Maryland Electron Ring (UMER) [21]. UMER is small compact electron storage ring with a low energy (10 keV) but relatively high beam current (1-100 mA). It is designed to study the physics of electron beams in both the emittance and space charge dominated regimes and to study physical processes that can be scaled to higher beam energies.

UMER is well-suited for the experimental study of halos, since (a) it can generate beam with and without halos easily; (b) has a number of phosphor imaging screens which can generate a substantial amount of light; and (c) the beam is highly stable and thus reproducible over many pulses. This allows us to acquire images with frame integration with minimum error due to shot-to shot variations. Table 2.1 lists the key parameters of UMER and Fig. 2.2 shows a schematic of the UMER layout.

Table 2.1: UMER design parameters [23]

| | |
|---|---|
| Beam Energy | 10 keV |
| $\beta = v/c$ | 0.2 |
| Pulse Length | 20-120 ns |
| Current | 0.5-100 mA |
| Ring Circumference | 11.52 m |
| Lap Time | 197 ns |
| Pulse Repetition Rate | 10-60 Hz |
| FODO Period | 0.32 m |
| Zero-current Phase Advance | 0.760 |

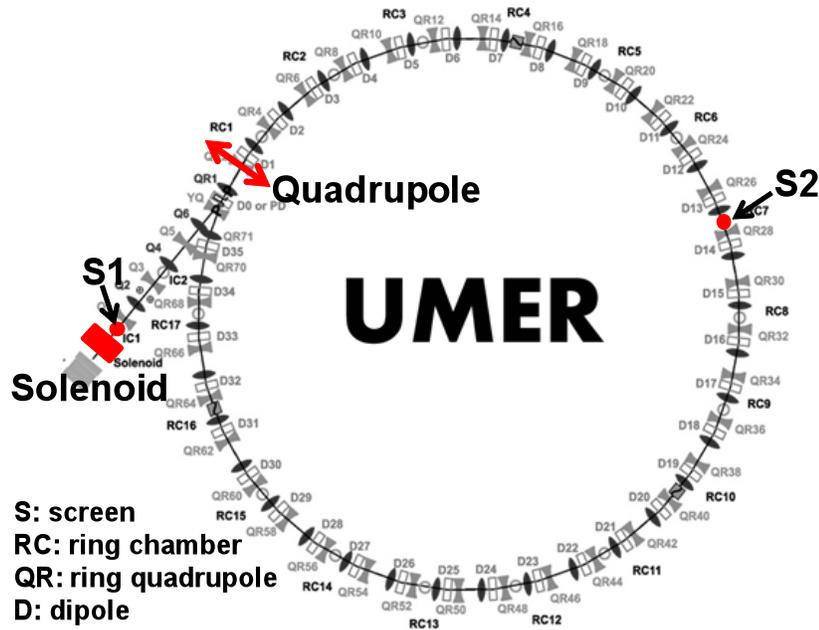

Figure 2.2: Schematic layout of UMER.

A key feature of UMER is its ability to generate beams with different intensities. The related intensity parameter $\chi$, the ratio between space charge force and external focusing force, can be varied to generate a beam, whose behavior is dominated by the emittance, to one where space charge dominates the dynamics. The intensity is varied by using apertures to change the beam current. A mechanically rotating wheel with a number of apertures is located right after the electron gun exit to do this. The variable intensity allows the study of halos in different types of beams, as well as the effects of perturbations such including magnet misalignment, beam mismatch, space charge on halo formation and evolution. Table 2.2 shows a list of aperture size, current, emittance and intensity parameter available at UMER [23].

Table 2.2: Beams in UMER [23]

| Aperture# | $r_0$ (mm) | $I$ (mA) | $\varepsilon$ (µm) | $\chi$ |
|---|---|---|---|---|
| 1 | 0.25 | 0.6 | 7.6 | 0.27 |
| 2 | 0.875 | 6 | 25.5 | 0.6 |
| 3 | 1.5 | 21 | 39.0 | 0.32 |
| 4 | 2.85 | 78 | 86.6 | 0.84 |
| 5 | 3.2 | 104 | 97.3 | 0.90 |

2.3 Optics Setup at UMER

The optical system for UMER shown in Fig 2.3 is essentially the same shown in Fig. 2.1 except that two additional mirrors are incorporated into the design to meet space constraints. The image source at UMER is a 31.75 mm diameter glass screen that is coated with P-43 phosphor ($Gd_2O_2S$: Tb). This phosphor has an emission peaked in the green (545 nm) and a response time of 1.6 µs. The phosphor screen is oriented at 90 degrees with respect to the beam direction and the light is directed out of a vacuum system by a front surfaced mirror. The location and properties of each optical component in the system are listed in Table 2.3.

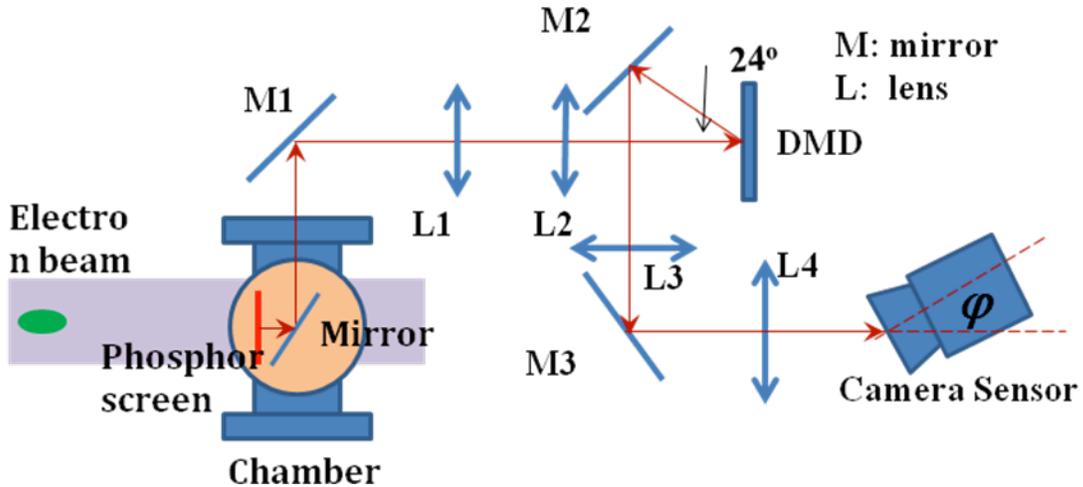

Figure 2.3: Sketch of optical system used at UMER.

Table 2.3: Parameters of UMER optical component

| Component | L*(mm) | Focal length(mm) | Diameter(mm) | Description |
|---|---|---|---|---|
| L1 | 600 | 200 | 75 | Achromat |
| L2 | 730 | 200 | 50 | Achromat |
| DMD | 826 | | | |
| M2 | 882 | | 50 | Front surface mirror |
| L3 | 974 | 100 | 50 | Achromat |
| M3 | 1054 | | 50 | Front surface mirror |
| L4 | 1094 | 200 | 50 | Achromat |
| CCD | 1186 | | | PIMAX2 |

*L is the distance from the phosphor screen

One of the design goals for the first channel of the optics is to achieve the best image resolution of the phosphor screen on the DMD. To accomplish this we adjust the magnification so that the image of the phosphor screen fills the shortest dimension of the DMD chip, which is a rectangle, 14.3 mm × 10.8 mm. Thus, the desired magnification for the first optical channel, $m$ =10.8 mm / 32 mm = 0.338. The use of two lenses in the first channel allows us to achieve this value and focus the image onto the DMD within the physical size constraints of our setup.

To simplify the analysis we consider each lens pair as a single effective lens. This is a good approximation when the distance between the two lenses is less than or close to the smallest focal length of either lens, which it is in our case. Results from a ray tracing code applied to our optical configuration confirm this approximation. A simplified diagram for the first channel is shown in Fig. 2.4.

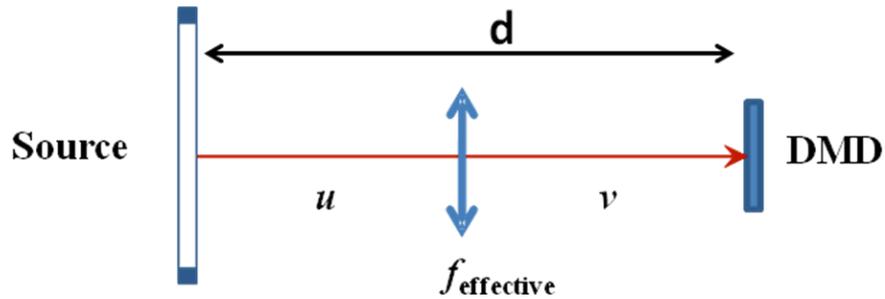

Figure 2.4: Schematic of the first optical channel.

Then using the magnification $m = u / v = 0.338$ and the total distance between the target and the DMD, $d = u + v = 826$ mm, where $u$ and $v$ are respectively the object and image distances from the effective lens. Using the lens equation: $1 / v + 1 / u = 1 / f_{\text{eff}}$,

$$f_{\text{eff}} = \frac{md}{(m+1)^2} , \tag{2.1}$$

so that $f_{\text{eff}} = 156$ mm for the first channel.

In practice, we choose readily available lenses, L1 and L2 and manually adjust them to achieve focus of the target onto the surface of the DMD. Since there is light scattering from the aluminum edge of the DMD chip holder that may affect image quality at the DMD boundary, we slightly reduce the image size on the surface of DMD to avoid this. Thus, we set the final magnification, $m = 0.270$, which is little smaller than the one calculated from Eqn. 2.1.

*Rotational Compensation*

As mentioned previously, the rotation axis of each micro-mirror is along the surface diagonal. Then, if the DMD is positioned so that the array's vertical axis is normal to the horizontal plane, the micro-mirrors on the chip will reflect the incident light out of the horizontal plane. In order to compensate for this, we rotated the DMD by 45° about the axis perpendicular to the plane of the micro mirrors in order to make the rotation axis of each micro-mirror coincide with the vertical as seen in Fig. 2.5. As a result, regardless of the orientation of the micro-mirrors, the light path is maintained in the horizontal plane. This means that the center of all the optical components can be set in the same plane. This greatly simplifies the positioning and the alignment of the optics in both channels and is an essential feature of the optics design.

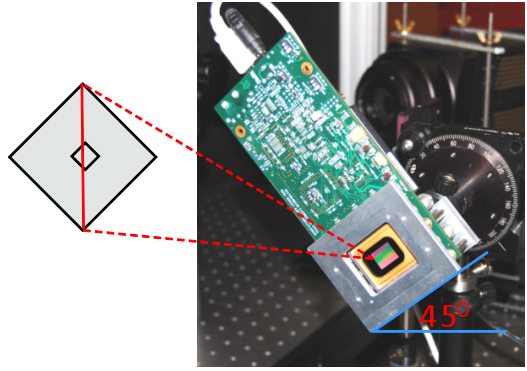

Figure 2.5: Picture of rotated DMD (Discovery 1100) and an enlarged sketch of a micro-mirror.

Similar to the first optics channel, the two lenses of the second channel allow a variable magnification and independent focus of the image of the DMD onto the camera, a 16 bit, variable gain, gated, cooled ICCD with 512 × 512, 13 µm square pixels (PIMAX2 manufactured by Princeton Instruments Inc.). This type of camera is not essential but its features are convenient for our optical system performance tests and halo imaging experiments. The transverse size of the camera's intensifier is 15.8 mm × 15.8 mm. It consists of a photocathode, micro-channel plate (MCP) and an output fluorescent screen. A tapered fiber optic bundle reduces the image on the output side of the intensifier to the size of the CCD (12.4 mm × 12.4 mm).

In order to have the best spatial resolution of the image on the CCD, we require that the image of the phosphor screen on the input side of the intensifier be as large as possible, but still allow a clear view of the edges of the DMD. The latter are used to calibrate the number of DMD pixels in terms of CCD pixels which is used in the mask generation algorithm discussed below. To achieve these requirements, we set the magnification, m = 1.003 using lenses L3 (100 mm) and L4 (200 mm). With this magnification and knowledge of the distance between the DMD and the CCD sensor ($d$ =

364 mm), Eq. 2.1 can be used to determine the effective focal length, $f_{eff}$ = 91 mm, of an idealized single lens which could be used to replace the pair L3, L4 of the second optical channel. This 'effective lens' will be used below in other calculations.

*Scheimpflug Compensation*

Note that for the second channel, the image of the secondary source, i.e. the image on the DMD produced by the first channel and viewed by L3 and L4 is not parallel to the effective lens plane but is inclined at 24° as is shown in Fig. 2.6. This is due to the fact that the micro-mirrors, which transmit the desired portion of the beam image from the first channel to the second channel, are flipped +12° when activated, so that incoming light rays from the first channel are reflected at twice this angle. Figure 2.6 illustrates this for rays parallel to the optical axis of the first channel.

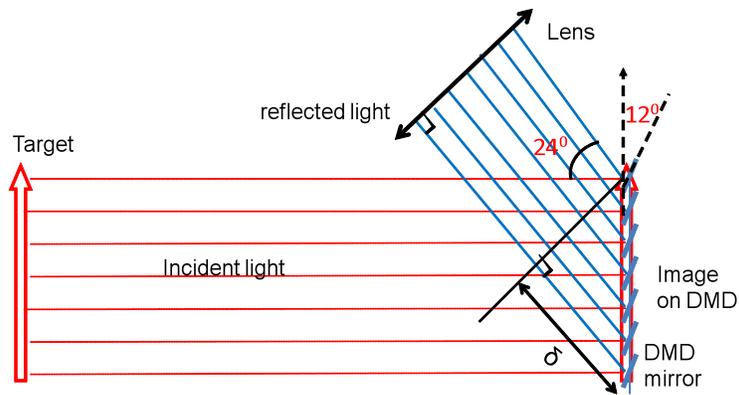

Figure 2.6: Path length difference generated by DMD tilted mirror.

The effective lens optical layout for the second channel is shown in Fig. 2.7, where line FDC is the light path of the central ray.

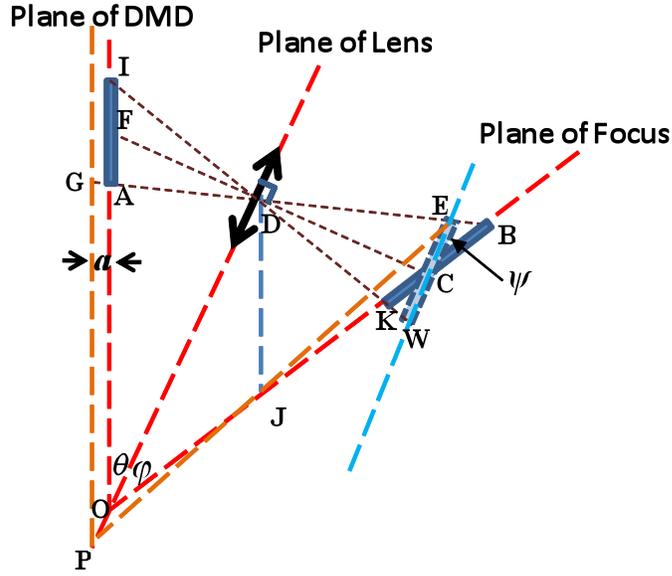

Figure 2.7: Scheimpflug Compensation diagram.

If we position the camera sensor normal to the optical axis *FC*, we will obtain an image with a non-uniform focus and magnification in the horizontal plane due to the difference in path lengths of the rays emanating from the DMD. To compensate for this, the CCD sensor must be tilted by an angle $\varphi$. This method is commonly used in photography and is known as Scheimpflug compensation [24] and the angle $\varphi$ is called the Scheimpflug angle. Based on the Scheimpflug principle, the planes of DMD, the effective lens and the CCD sensor coincide at one point indicated as point *O* in Fig. 2.7. The Scheimpflug angle can be simply determined from Fig. 2.7:

$$\varphi = \arctan(\frac{v}{u}\tan\theta) \tag{2.2}$$

where *u* (*v*) is the distance between the effective lens and the object (image), indicated as line *FD* (*CD*) in Fig. 2.7, where $\theta = 24°$. For the second channel, $u + v = d = 364$ mm, and $u / v = m_c = 1.033$, so *u* and *v* are 185 mm and 179 mm from Eqn. 2.1. Thus the Scheimpflug angle for the second optical channel is 24.7°.

*Depth of Field*

An analysis of the Scheimpflug condition shows that if the source plane, in this case the plane of the DMD, is moved a distance *a*, the intersection point *O* moves to point *P* and the plane of focus is rotated around the point *J*, which is known as the 'hinge point', such that the line *EP* lies in a new plane of focus. When Scheimpflug compensation is applied to view-camera photography, the camera system is designed so that film plane can be physically rotated about the hinge point. However, in our experiments this is not convenient and, instead, we rotate the CCD sensor about *its center* in order to achieve the best focus for the second channel. In doing so we empirically find that the best overall focus is achieved at the angle, $\varphi=24°$ (see line *ECW* in Fig. 2.7), which is $\psi=0.7°$ less than the value calculated (24.7°) from Eq.2.2.

In order to see the effect of this difference consider the following. In Fig.2.7, the line segment *FA* is half size of the DMD (indicated as the line segment *AI*) and the point *F* is the center point. The best focused image of the DMD will lie on a plane containing the line segment *BK*. However, if we rotate the camera by an angle $\psi$, the image on CCD will fall on *EW* instead of *BK*. According to hinge rule [24], if we connect any point in line segment *CEW* with hinge point *J* to define a new plane of focus, i.e. the plane containing *EJP*, and track it back, the source plane corresponding to the new plane of focus will be parallel to the plane of the DMD, i.e. the plane containing *GP*. Therefore, there is a distance between the plane of DMD and the corresponding source plane for each point on *CEW*. Because we rotate the CCD sensor about its center, the corresponding plane for *F* is the plane of the DMD and this distance is zero. Since *E* is the point that is out of focus the most, the corresponding plane containing *GP* is the maximum distance from the plane of the DMD. We can regard the distance *a* as the depth of field, as shown in Fig. 2.7. The depth of field is related to the object distance *u*, the image distance *v*, the size of the DMD and the difference in angle between the calculated Scheimpflug angle and the empirical angle, $\psi$ which can be determined from the geometry shown in Fig. 2.7. For our optical system, the size of DMD is 10.8 mm and $\psi$ = 0.7°, so that *a* = 0.06 mm, i.e. the maximum depth of field is very small compared with the smallest dimension of the DMD.

*Distortion*

When Scheimpflug compensation is applied to the horizontal plane, there will be an unavoidable distortion due to the difference in magnifications in both the horizontal and vertical directions.

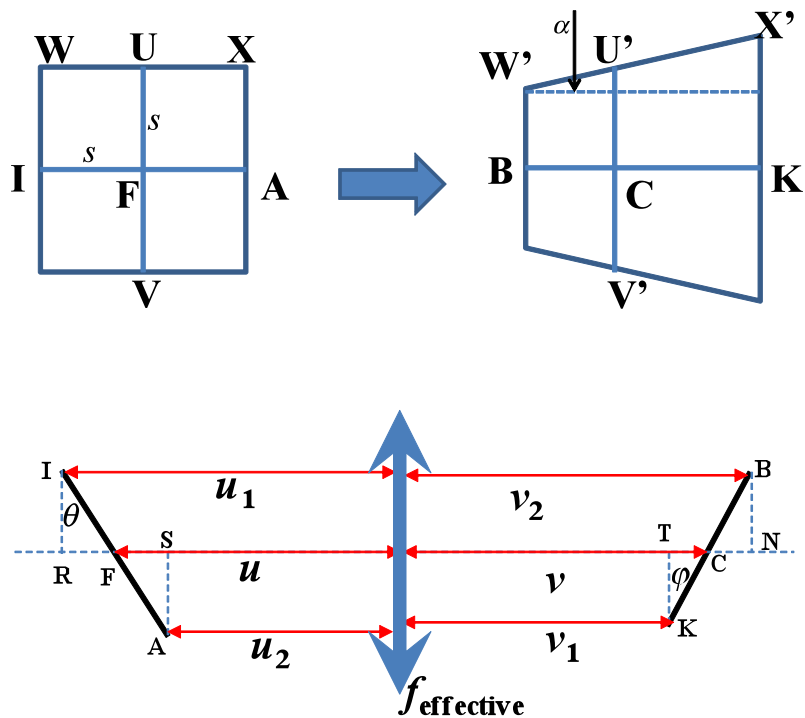

Figure 2.8. top: the effect of distortion due to Scheimpflug compensation; bottom: diagram for calculation.

As shown in Fig. 2.8 (a), a square object with half size *s*, will transform to a trapezoid on the image plane. Because the vertical line *UV* is perpendicular to the optical axis, the segment *UF* and *FV* will have the same magnification. However, for the horizontal line which is tilted by an angle $\theta = 24°$, as shown in Fig. 2.8 (b), the segments *IF* and *AF* will have different magnifications when they are imaged to lines *BC* and *KC*, respectively. Moreover, the off-axis vertical segment *WI* and *XA* will have different magnifications when imaged to *W'B* and *X'K*. It is easy to verify that the ratio *KC/BC* is equal to *X'K /W'B*. We define this ratio as the distortion *D*:

$$D = \frac{KC}{BC} = \frac{X'K}{W'B} \quad . \tag{2.3}$$

From Fig. 2.8 (b),

$$BC(KC) = \frac{BN(KT)}{\cos\varphi} = s\, m_{RC(ST)} \frac{\cos\theta}{\cos\varphi} \quad , \tag{2.4}$$

where $m_{RC}$ and $m_{ST}$ are the magnifications of line segments *IR* and *AS* respectively and:

$$m_{RC(ST)} = \frac{f_{eff}}{u_{1(2)} - f_{eff}} = \frac{f_{eff}}{u \pm s\sin\theta - f_{eff}} \quad , \tag{2.5}$$

where $u_{1(2)}$, as shown in Fig. 2.8, are the source distances from the effective lens, and the subscripts 1, 2 refer to line segments *IR, AS* respectively. Finally, we can express *D* as

$$D = \frac{m_{RC}}{m_{ST}} = \frac{u - s\sin\theta - f_{eff}}{u + s\sin\theta - f_{eff}} = \frac{\dfrac{f_{eff}}{m_C s} - \sin\theta}{\dfrac{f_{eff}}{m_C s} + \sin\theta} \quad , \tag{2.6}$$

where $m_c$ is the magnification for the second channel. Eq. 2.6 shows that the distortion is significant only when the DMD size is magnified to a value comparable to the effective focal length, which is *not* the case for our system.

The trapezoid angle $\alpha$, can be determined from the geometry and

$$\tan\alpha = \frac{s\sin\theta\cos\theta}{\sqrt{u^2\cos^2\theta - 2uf_{eff}\cos\theta + f_{eff}^2}} \qquad (2.7)$$

It can be easily verified that when the tilted angle $\theta = 0°$, the distortion $D = 1$ and trapezoid angle $\alpha = 0$.

For the UMER optics, $u = 179$ mm, $v = 185$ mm, $s = 5.4$mm and $f_{eff} = 91$ mm, and $\theta = 24°$, so that the distortion $D = 1.052$, and the trapezoid angle $\alpha = 0.45°$; therefore the distortion effect is negligible. This is verified by direct observation of a test target used in our bench tests, which consists of a gridded circle, whose diameter (32 mm) is the same as the phosphor screen used in our halo measurements (see Fig. 2.9).

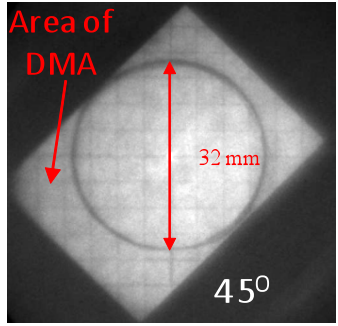

Fig. 2.9: Image of a test target on the ICCD camera.

2.4 Mask generation algorithm

In order to generate a mask for the beam core, a coordinate transformation and rescaling is necessary, because of the 45° orientation of DMD and the different number of pixels in DMD and the CCD sensor. The algorithm is shown schematically in Fig 2.10.

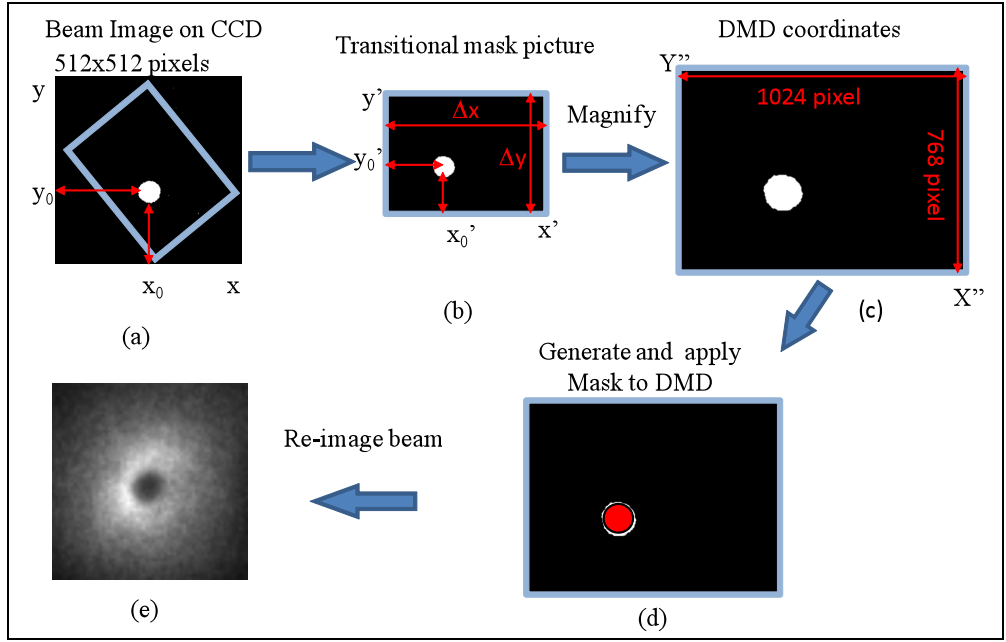

Figure 2.10: Schematic of mask generation algorithm.

The procedure is as follows. First we take a calibration picture to view the edges of DMD chip on the CCD, and thus determine the DMD chip size ($\Delta x$, $\Delta y$) as well as the equations describing the edges of DMD chip in term of the CCD coordinates $(x, y)$ as shown in Fig.2.10 (a). Then, we generate a single bit, transitional picture, $\Delta x * \Delta y$ in size, with all pixels 'on' set to 0 and all pixels 'off' set to 1. After we take a beam picture, any point of interest $(x_0, y_0)$, can then be transformed to the transitional picture as in terms of coordinates $(x', y')$ by calculating the distance between the point and the edges of the DMD. We then choose selected discrete points on the beam image in this manner and set the pixel intensity value to 1 in order to generate a mask, which will (reject), i.e. block the beam core shown as the red area in Fig. 2.10 (b).

There are several ways to define the mask. For example, we can select points in the core, by 1) specifying a particular geometric area (e.g. a circular disk) visually to define the core, or 2) setting a chosen intensity threshold value to define the core. Note that the area (Δx * Δy) is still specified in units of the pixels of the CCD, which is much smaller than the size of DMD. Thus, for next step as shown in Fig. 2.10 (c), we linearly magnify the transitional picture by the ratio Δx*Δy divided by the area of the DMD chip and then generate the final 1-bit mask.

We have written a MATLAB code to generate the mask. The code input is an image of the beam recorded by the PIMAX2 camera control software (Winview32). Using the algorithm described above, the code generates a single bit 1024 x 768 mask picture file in a bitmap (.bmp) format with black pixels set, e.g. by a threshold intensity level in the input beam image to define the mask, and white pixels, set by all intensity levels less than the threshold, thus represent the unmasked region. This '.bmp' file can be read by the software supplied with the Discovery 1100 DMD and the 4100 DMD. The software first clears the current status of the micro-mirrors and then sets them to either +12° or -12°, corresponding to black (0) or white (1). Once the mask is applied to the DMD, the masked beam is reimaged on the ICCD camera and a final image is generated by integrating by the number of beam pulses required to bring the peak intensity of the final image close to the saturation level of the camera sensor. The gate feature of the PIMAX2 is utilized to do this by setting the gate width to slightly exceed the UMER beam pulse width (100ns) and accumulating images for a set number of gates corresponding to the necessary number of beam pulses. This feature minimizes the background from stray light sources.

## 3. Experiment Results

3.1 Optical bench tests

*DMD Diffraction Effects*

As noted above the DMD behaves like a 2D optical grating. If illuminated by a single wavelength laser source, a cross like diffraction pattern similar to that of a rectangular mesh will be observed. When all the micro-mirrors are rotated by +12° the DMD becomes a blazed grating with the central order reflected in the direction +24° in the horizontal plane with respect to the incident laser beam. When the DMD is rotated by 45° the diffraction pattern also correspondingly rotates; the resulting pattern is shown in Fig. 3.1 (a). Note that the central order has been suppressed so as not to saturate the imager. When a uniform source of *white light* illuminates this "blazed" grating, the light is further dispersed in the horizontal plane producing the Fraunhofer pattern shown in Fig. 3.3 (b). These pictures were obtained by imaging the light diffracted by from the DMD (all pixels 'on') in the focal plane of a 200mm focal length lens.

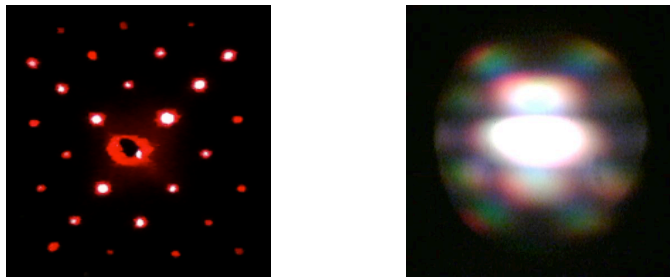

Figure 3.1: (a) Single wavelength diffraction pattern; and (b) white light diffraction pattern, both formed from a $45^0$ rotated DMD with all pixels set at $+12^0$.

Note that for the white light diffraction pattern all the orders are smeared in the horizontal plane. This is due to the effect of both wavelength dispersion and overlapping of the light from the central spot and nearby first order diffraction spots. The latter effect is particularly evident in the central and first orders. The picture further makes it clear that most of the light reflected by the DMD is contained in these two orders.

If the incident light on any grating is non-uniform but has a known distribution, i.e. is an 'object', any order of the light diffracted by the grating can be used to image the 'object'. An aperture in the optics will filter out higher order diffraction spots which ultimately will reduce the imaging resolution, so this must be checked for a given application. However, if the numerical aperture of the second channel optics is sufficient to accept the central two orders, most of the light diffracted by the DMD will be relayed into the second channel. This is indeed the case for our optics. We have traced the rays corresponding to the angles of the central and first order diffraction spots with an optics code to insure that there is no vignetting of these rays through the optics of channel two.

*Spatial Resolution*

We measured the spatial resolution of our optical system by imaging a "knife edge" resolution target, constructed from a rectangular piece of black anodized aluminum foil (Cinefoil) mounted on a white card, onto the DMD. The card is backlit with an adjustable intensity 'white light' source (i.e. an incandescent lamp). In these bench tests we make use of a higher resolution PIMAX2 camera with a 1024 × 1024 pixels CCD array. Each pixel of the CCD sensor is 13 µm × 13 µm.

To insure that the DMD plane is in good focus on the camera to begin with, we program the DMD to accept a well known test image, i.e. a black and white checkerboard, which is included in the software supplied by the manufacturer to control the DMD. This pattern is ideal for adjusting the focus of the second channel since it originates with a few microns of the surface plane of the DMD chip and has multiple sharp black-white boundaries, i.e. the checkers. Once this source is focused onto the CCD, we turn all pixels to the 'on' state to reflect the resolution target image from the first channel into the camera light path. We then adjust the focus of the first lens system L1 and L2 to produce the best focus of the resolution target on the camera, without moving lenses L3 or L4. Fig. 3.2 shows two views of the resolution target; the left hand side is a full view of the entire target, the right hand side a magnified view of the corner of the black rectangle portion of the target.

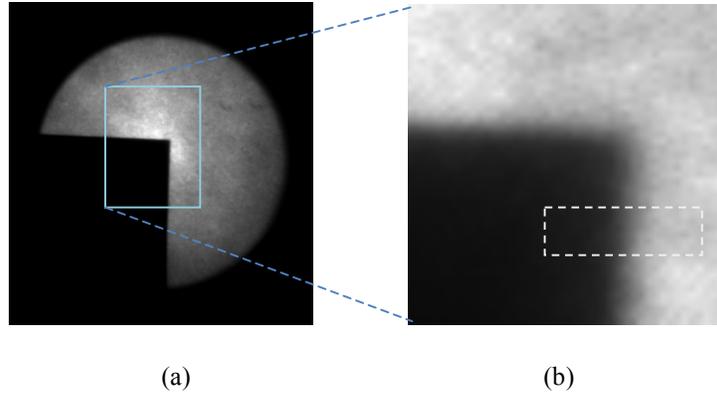

(a) (b)
Figure 3.2: (a) Resolution target; left: full view, right: (b) expanded (× 16) view.

Vertical and horizontal line scans of the corner of the black corner of the target at the pixel level show an 'S' shaped dispersion curve. Fig. 3.3 shows a horizontal scan which is averaged over 4 vertical pixels (see the dotted white box shown in Fig. 3.2 (b)). To analyze the *resolution* of the image we assume that the point spread function can be represented by a 2D Gaussian and convolve this function with the source intensity distribution $A(x,y)$. An line scan across the image, e.g. $I(x)$, the brightness along the linear scan ($X$ direction) normal to the sharp linear boundary ($Y$ direction, at $Y = 0$) is described by the convolution integral:

$$I(x) \propto \iint A(x', y') \exp\left(-\frac{(x-x')^2}{2\sigma_x^2} - \frac{y'^2}{2\sigma_y^2}\right) dx' dy' \qquad (3.1)$$

where we assume A(x,y) = const. at $X > 0$ and $A(x,y) = 0$ at $X < 0$. A similar expression can be derived for a Y scan. One can easily show that the resulting intensity scans are error functions. Simple fits of the experimental horizontal and vertical scan data to the corresponding error functions and their respective Gaussians functions, which are the derivatives of these functions, is shown in Figure 3.3 (b). The widths of the horizontal and vertical Gaussians are 3.0 and 2.1 respectively. The width provides as estimate of the resolution of the second optical channel.

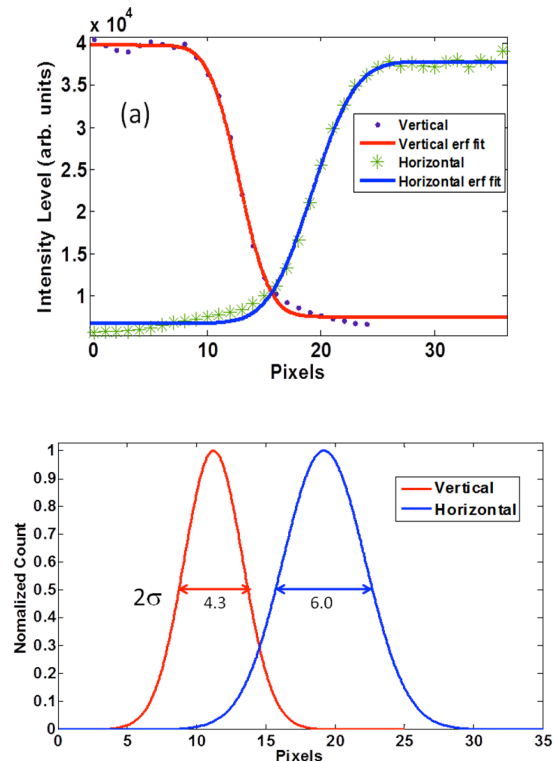

Figure 3.3: (a) Horizontal and vertical scan of resolution target corner shown above in Fig. 3.4 and (b) related Gaussian functions.

We compare these measured values of the resolution of the DMD optical system, i.e. the second channel, to that of a 'standard' optical system in which the DMD is replaced by a simple mirror. To perform this measurement we use a thin (1mm) front surface aluminized silicon mirror, which is temporarily affixed onto the aluminum frame housing of the DMD and stands off from the surface of the DMD by about 0.5 cm. We slightly adjust the angle of the mirror and position of the lenses L1 and L2 to view the resolution target with the same magnification as the optical system which included the DMD. The major difference in the optics when the mirror is used is that Scheimpflug compensation is not required. So for the resolution test with the mirror we rotate the CCD camera back to its initial orientation, i.e. perpendicular to the optical axis of the second optical channel. We then observe the knife edge resolution target and follow same procedure described above to measure the resolution from the knife edge target image line scans. The measured width of the black-white transition region in both the horizontal and vertical directions is $\sigma \sim 3$ pixels, which is approximately the same value measured for the DMD optical system. This indicates that the optical resolution of the DMD optics is essentially the same as that of the 'standard' mirror system.

*Single Pixel Response*

We have also measured the single pixel response of the second optical channel. To accomplish this, a uniform white light source beam is directed onto the entire DMD but with only a single micro-mirror activated. The light from this single illuminated pixel is imaged onto the CCD via lenses L3 and L4. Fig. 3.4 shows the resulting image (shown as a negative) as well as horizontal and vertical line scans across the image. The scans show that the width of a Gaussian fit to distribution of intensity in both the horizontal and vertical directions is $\sigma \sim 3$ pixels, which, interestingly, is the same as the measured optical resolution of the entire optical system. This result means that the DMD does not significantly influence the resolution or PSF of the optics system. Furthermore, since the contribution to the PSF of the second channel is comparable to that of the overall system, the width of the PSF of the first channel must be less than that of the second channel or very close to it.

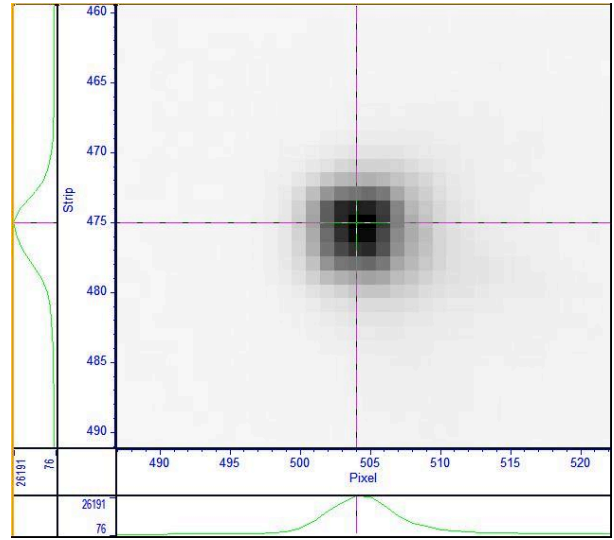

Figure 3.4: Response of a single DMD pixel to uniform white light illumination.

*High Dynamic Range Measurement of the PSF of the Primary Optical System*

As far as dynamic range of the optical imaging system is concerned, it is only necessary to measure the PSF of the first channel. The reason for this is that the second optical channel merely reimages the first image from the DMD plane with or without a mask in place, with an inherently low dynamic range imaging system.

The optical masking technique we will describe below uses a series of images each of which are taken using a CCD camera with a (DR ~100-1000). Thus the wings of the PSF of second channel below $10^{-2}$ or $10^{-3}$ of the peak intensity of any masked or unmasked image of the beam are not visible to the CCD camera. Any contamination of the true beam halo due to the wings of the real PSF will be visible in the first channel and affect image on the DMD. This means that if the wings of the PSF exceed the beam halo in first image they will also exceed the beam halo in the distribution, which is reconstructed using the DMD masking method. Similarly, if the wings of PSF are below the true beam halo level, this will be the case in the reconstructed image as well. Thus, for all intensive purposes it is sufficient to measure the PSF with high dynamic range for the first channel only.

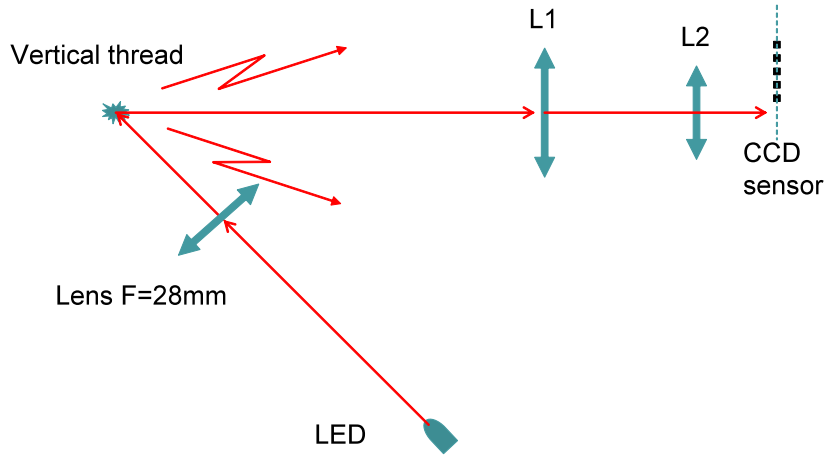

Figure 3.5: Sketch of the experimental setup for PSF measurement of the first optics channel.

In order to do this, we use a wide band (white), 'point like' source with a homogeneous angular distribution, which closely mimics that of the phosphor screen used at UMER. A schematic of measurement system is shown in Fig. 3.5. The source is a white thread illuminated by light produced by a light-emitting diode (LED) that is 4 mm in diameter. The LED light is focused onto the thread by a Nikon camera lens which has focal length $f = 28$ mm and is oriented so that the normal input aperture of the lens faces the LED. The thread is a very good diffuser and scatters the focused LED light uniformly into a wide angle. We verified this by imaging the angular distribution with a CCD camera placed in the Fourier plane of the first optical channel and confirmed that the irradiance across the sensor was uniform. The image of the LED on the thread is 0.45 mm in length and 0.15 mm in width, which corresponds to the diameter of the thread. When the light scattered by the thread is focused onto the CCD sensor by the first optical channel (i.e. lenses L1 and L2) the size of geometrical spot is on the CCD sensor is 0.04 × 0.12 mm, or about 1.5 × 4 pixels.

We measured the PSF of the first channel by progressively shifting the bright central spot in the image of our source away from the active area of sensor of CCD camera via a mechanical linear actuator and successively applying neutral density filters to attenuate the light to avoid saturating the CCD. By means of this technique we were able to utilize the whole size of the CCD sensor and achieve a dynamic range DR~$10^7$. The results are shown in Fig. 3.6.

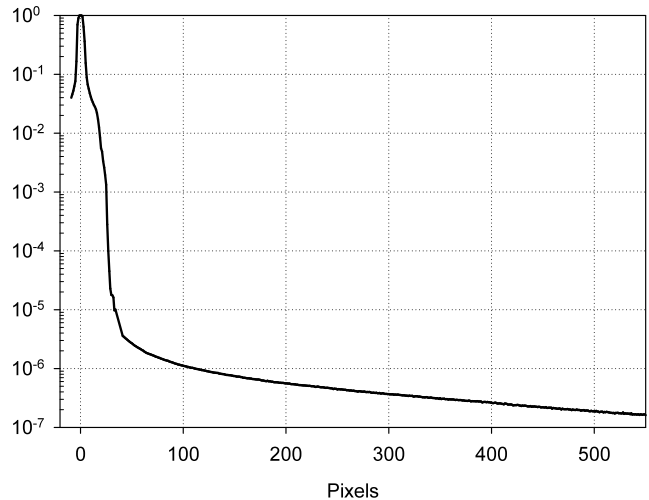

Figure 3.6: Measured PSF of the first optical channel

We note that the measured PSF has a much wider FWHM than the PSF, calculated from diffraction theory, i.e. the Airy disk. Furthermore, the intensity of the wings of the measured PSF in the interval 100-500 pixels is several orders of magnitude greater than the level calculated from diffraction theory. This is due to the combined effects of scattering, aberration, non-uniformities and diffraction in all the elements in the optical transport. The importance of this measurement will become apparent later when we discuss the dynamic range measurements done using the UMER electron beam.

Also, because of the finite size of our source, the measured PSF differs from that of a 'true point source', especially in the region close to the source. However, at large distances from the source the intensity distribution will be close to that of the 'true PSF' because it is not affected by the size of the source at sufficiently large distances, i.e. when ratio of distance over the source size is large.

3.2 Beam Imaging Studies at UMER IC1

*Halo Generation*

Tests of the performance of the optical system with a real beam are done on UMER at diagnostic chamber IC1 (the first injection chamber), shown as 'S1' in Fig. 2.2, where the beam core is approximately round. The electron gun of UMER [25] has a circular thermionic cathode source with a radius of 4 mm. The anode-to-cathode distance is 25 mm. The grid is positioned between them, 0.15 mm away from the cathode. The grid is biased with a negative voltage with respect to the cathode, to avoid the flow of electrons into the injector, until a positive over-riding beam pulse is applied. The length and repetition rate of the beam pulse is described in Table 2.1. During the high level of the beam pulse, the beam current is allowed to flow into the injector of the ring.

In normal operation we set the bias voltage to -30V. At this setting the beam current pulse has flat top shape with little fluctuation and the beam is round with almost no halo. However, when we alter the bias voltage to -50 V, we observe that the beam current fluctuates with a large variation, which indicates that each longitudinal slice of the beam pulse has a different transverse size. When all the slices are integrated a halo is observed. We use the resulting beam to: 1) test the filtering effectiveness of the DMD; 2) test the imaging quality of the optics; and 3) measure the dynamic range of the entire optical system, which is accomplished by changing the solenoid strength to focus the beam to a spot size on the phosphor screen S1, which is as small as possible.

*Extinction Properties of DMD*

We tested the effectiveness of the DMD to reject light away from the camera optical path (i.e. channel 2) in the following way. We first set all the DMD pixels to the "on" state, so the whole image of the beam is transported to the CCD sensor. This image is shown in Fig. 3.7 (a). The peak intensity is about 61,500 counts, which is near the maximum intensity level (65,355) of the CCD pixels; above this level, the CCD pixels are saturated. The camera was set to integrate 180 beam pulses to reach this level. Next, we set all the pixels of DMD to the "off" state. In order to compare with the previous one, we also integrated over 180 beam pulses resulting in the image shown in Fig. 3.7 (b). Comparing these two images in the beam region, the image with the all DMD pixels "off" has an intensity level at least $10^3$ less than the same region with the pixels "on". Fig. 3.7 (b) shows that in the beam region level (< 50 counts) is almost at the noise level. Notice that there are two visible lines (but with very small number counts ~$10^2$) seen in the image shown in Fig 3.7 (b). These lines are due to stray light scattered from the edges of the DMD chip.

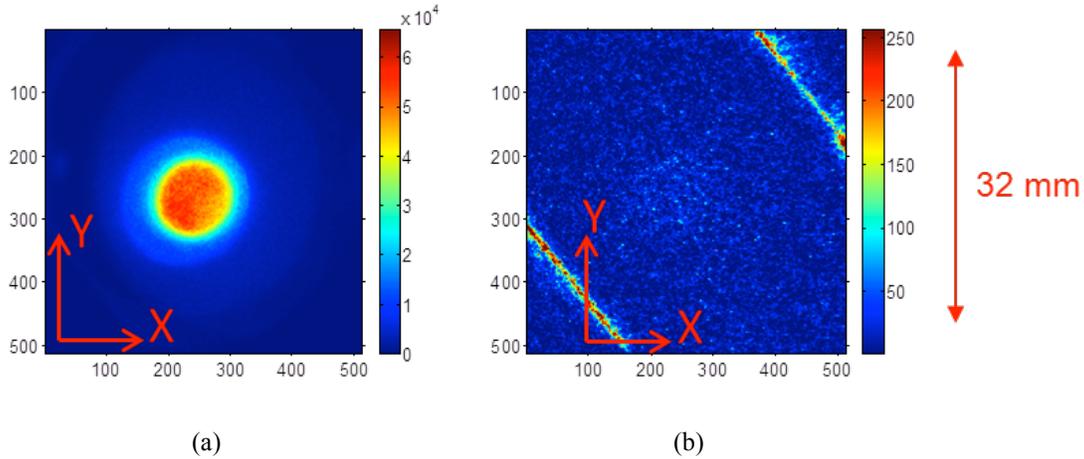

(a)                               (b)

Figure 3.7: (a) Image with all DMD pixels set to $+12^0$; (b) image with all pixels set to $-12^0$; vertical and horizontal coordinates are given in terms of CCD pixels

*Image Quality Tests*

Secondly, we did an experiment to test if the DMD affected the final image quality. Three different conditions are used as showed in Fig. 3.8: (a) is in the normal condition where all the DMD chip pixels are set to the "on" state and Scheimpflug compensation is used; b) is power off condition where all the DMD pixels are in the floating state (no tilting) and no compensation is used; and c) the DMD is again replaced by a simple mirror and no compensation is used.

Fig. 3.8 shows the beam images taken for each of these cases. Note that the optics must be slightly adjusted between three configurations. This causes slight differences in the magnification and number of peak counts for each case. As a result, we normalize the intensities by the peak values in each image in order to concentrate on differences in the beam picture. Fig. 3.8 and Fig. 3.9 show that the beam shape as well as the horizontal and vertical line scans of the beam in all three cases are essentially the same. Note that the image is shifted due to the change of the configuration of the optics. From this data, we conclude that diffraction and scattering by the DMD mirrors has little if any effect on the quality of the beam images; the activated DMD with Scheimpflug compensation produces as high a quality image as that of the optical system that uses a simple or 'flat' segmented mirror, i.e. when all the micro-mirrors of the DMD are in the power off or floating state.

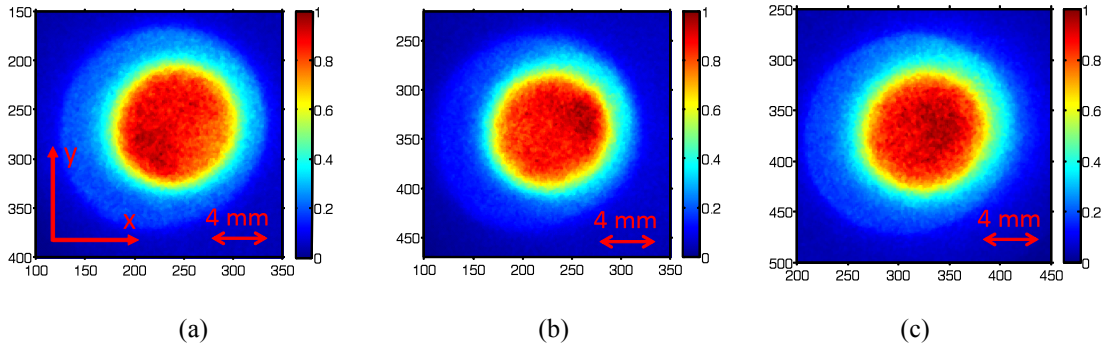

Figure 3.8: Comparison of beam images with: (a) all DMD pixels set to '+' and Scheimflug compensation; (b) all DMD pixels floating and no compensation; (c) simple mirror and no compensation.

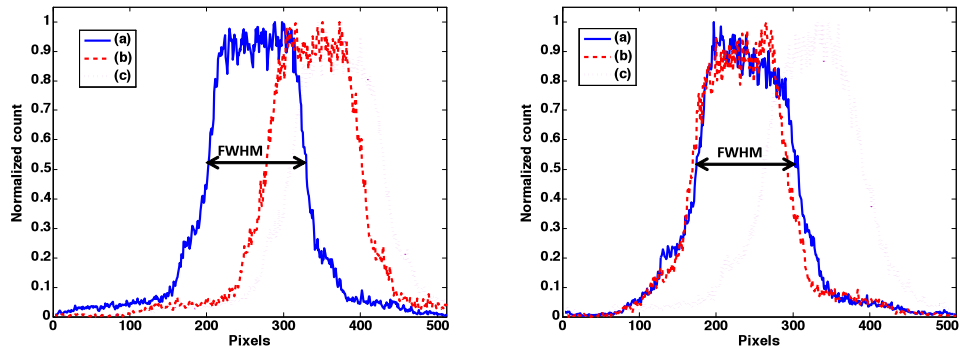

Figure 3.9: Left: Horizontal and right: vertical line scans of the three cases in shown in Figure 3.6; the FWHMs are 129, 124,129 pixels respectively pixels for the horizontal scan and 128, 122, 123 pixels respectively for the vertical scan.

*Dynamic Range of the Optical System*

In order to determine the dynamic range of the entire UMER optical system, we observed an intense beam (I=21mA) focused to a 2.85 mm diameter spot on the phosphor screen, which is the highest current and smallest size that can be achieved with our solenoid focusing magnet. We then generated a number of circular masks with different radii but a common origin, i.e. the position of the peak intensity of the beam observed without a mask. For each mask, we integrated over the appropriate number of frames (the number written underneath each photo of Fig. 3.10) that was necessary to bring the peak intensity in the image close to the saturation level of the camera. Note the small highlights visible in the upper left hand part of pictures number 3 and 4. These are due to scattering of the phosphor light from the metal edge of the screen. To block out these undesirable highlights we created additional small masks on the DMD, which are seen as black dots on the images number 5 and 6 of Fig. 3.10.

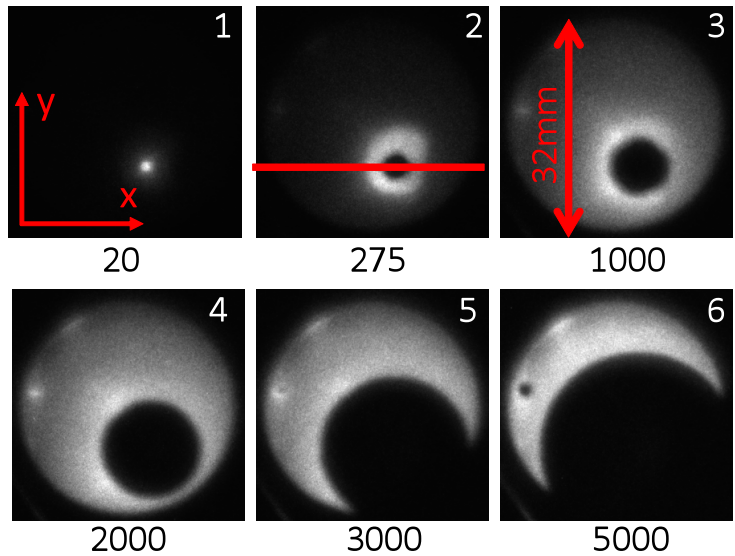

Figure 3.10: Images of the focused beam with the DMD set to concentric circular mask of successively larger radii; number below image denotes the number of frame integrations on the ICCD camera

To obtain a background image, we turned the beam off and integrated for the same number of frames that was used to obtain the beam image. Background subtracted images are shown in Fig. 3.10. By taking the horizontal line's scans of all beam profiles (note for reference the horizontal red line in Fig. 3.10) and normalizing by the number of frames of each image, we can present the results as a series of normalized plots as shown in Fig. 3.11.

Note the intensity fluctuations in tails of the beam profiles in Figure 3.11 decrease as more integration is applied. Moreover, the noise level outside the screen decreases to ~$10^{-5}$. Comparison of the recorded electron beam distribution with the measured PSF of the first optical channel (the blue dashed line in Fig. 3.11 (a)) makes it clear that the PSF "wings" due to the central, brightest portion of the beam (i.e. the core) do not obscure (i.e. over-illuminate) the halo region of the beam. Thus for our experimental conditions it was not necessary to use special Lyot stops or highly polished lenses to improve the PSF, as is done in beam coronagraphy, to achieve a system DR ~ $10^5$.

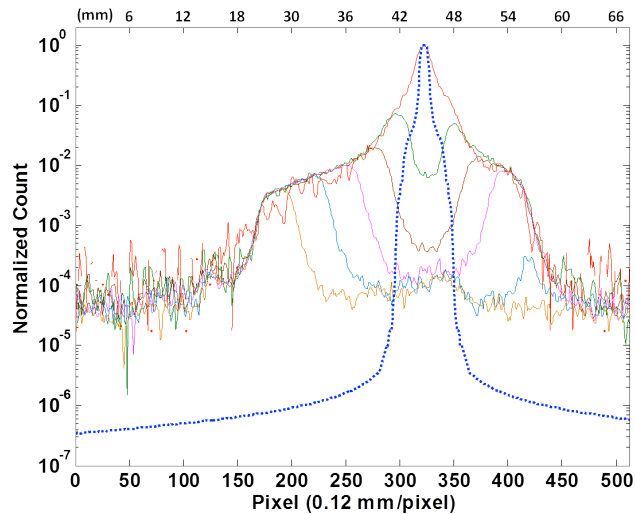

Figure 3.11: Normalized horizontal scans of selected beam images of UMER presented in Fig. 3.7; with the measured PSF shown in Figure 3.2 (blue dashed line ).

Using the data presented in Fig. 3. 10 and normalizing by the number of frames we reconstructed a high dynamic range beam image as shown in Fig. 3.12. For this reconstruction, each point in the Figure is chosen from an intensity interval, corresponding to the peak value to 1/60 of the peak value, for each image presented in Fig. 3.10. Then the value of the intensity for each pixel is normalized and converted into standard bitmap format using the relation $51*(5 + \log (I / I_0))$, where $I_0$ is the absolute maximum value of all the intensities.

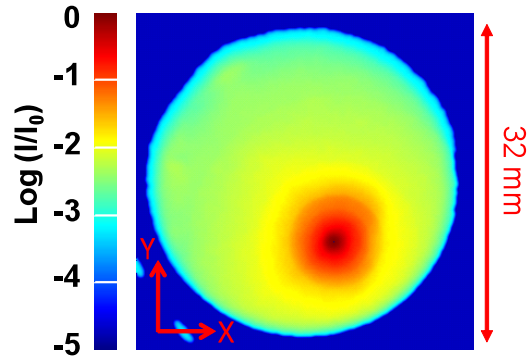

Figure 3.12: Reconstructed image showing colorized logarithmic intensity contours of the beam, phosphor screen periphery and background.

We note that our DR measurement at UMER is limited by the minimum size of the beam that can be obtained with our solenoid magnet, the maximum current density of the UMER beam, the saturation level of the phosphor and the size of the screen, i.e. not by the optics or the inherent properties of the DMD itself. Also note that the beam is off centered; this is due to the misalignment of the solenoid.

3.2 Beam Halo Measurements at UMER RC7

At RC7 (the 7$^{th}$ ring chamber), denoted by Screen 2 in Fig. 2.2, where the beam distribution is typically not round, we performed experiments to test the adaptability of the masking method to changes in the shape of the beam. In the UMER ring, the beam is not perfect matched, so the time integrated beam distributions we observe are complex. By varying the quadrupole upstream, for example QR2 (the 2$^{nd}$ ring quadrupole), we can empirically perturb the beam and thereby change the eccentricity of the beam core as well as the halo structure. Under these conditions we generated masks on the DMD that adapt to variations in the core of the electron beam. The results for two different beams with currents of 21 mA and 6 mA are shown in Fig. 3.13 and Fig. 3.14, respectively.

Fig. 3.13 shows a group of beam images taken with the 21 mA beam, where the numbers in the lower left are the number of frames integrated and the numbers between rows (a) and (b) is the threshold intensity levels used to generate the mask. We decreased the original quadrupole current strength ($I_Q$) by 12.4% and 28.8% to affect changes in both the beam core and the halo. A decrease in the quadrupole current results in a linear drop in the magnetic field and thus an increase in the focal length of the quadrupole.

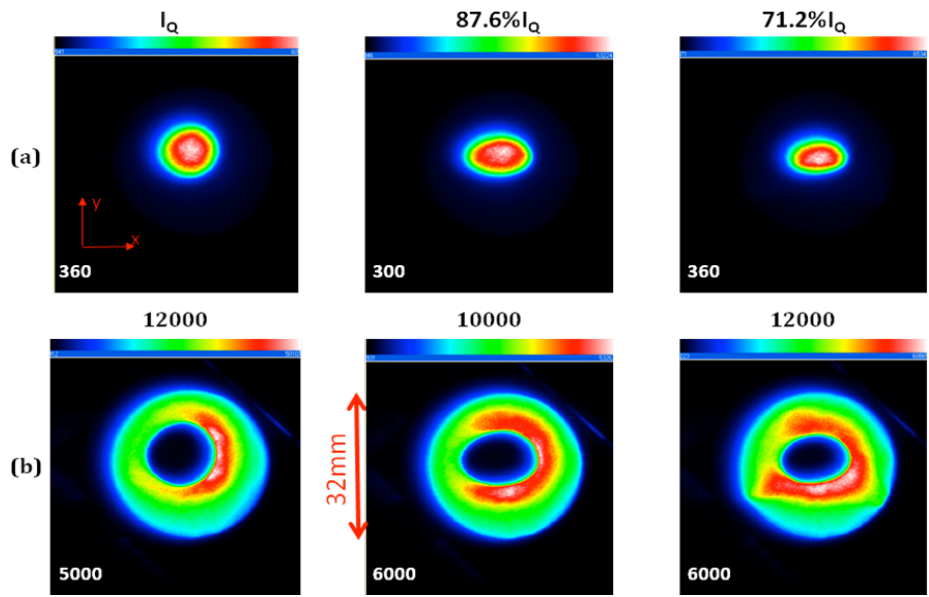

Figure 3.13: Comparison of core and halo with changing quadrupole strength
(a) Unmasked beam picture; (b) masked beam picture.

Row (a) in Fig. 3.13 shows photos of the beams for three current settings, which indicate that the core gradually enlarges in the *x* direction and shrinks in *y* direction and thus becomes elliptical. Row (b) shows the same beam after applying threshold masks which conform to the variations in the core shapes shown in row (a). We also see that decreasing the quadrupole current strength also affects the halo distribution. Notice that more particles are driven out of the core, as shown in Fig. 3.13, as the size of the high intensity and halo regions become larger.

In order to see the full extent of the halo, a much smaller beam with 6 mA current (radius = 0.875 mm) was used. We again decreased the original quadrupole current strength ($I_Q$) by 17.1%, 33.7% and 50.3% to see the effect on the beam core and halo. Fig. 3.14 shows the results. Comparing the pictures in row (a), we see the beam centroid gradually moves toward the negative *y*-direction. This may be the result of quadrupole misalignment with respect to the center of beam pipe or beam misalignment. Again as the quadrupole current decreases, the beam shrinks in the *x* direction and expands in the *y* direction, and as shown in row (b), as the quadrupole current decreases, particles escape from the beam core and appear to rotate in the halo region. The typical halo size is 2 or 3 times greater than that of the beam core.

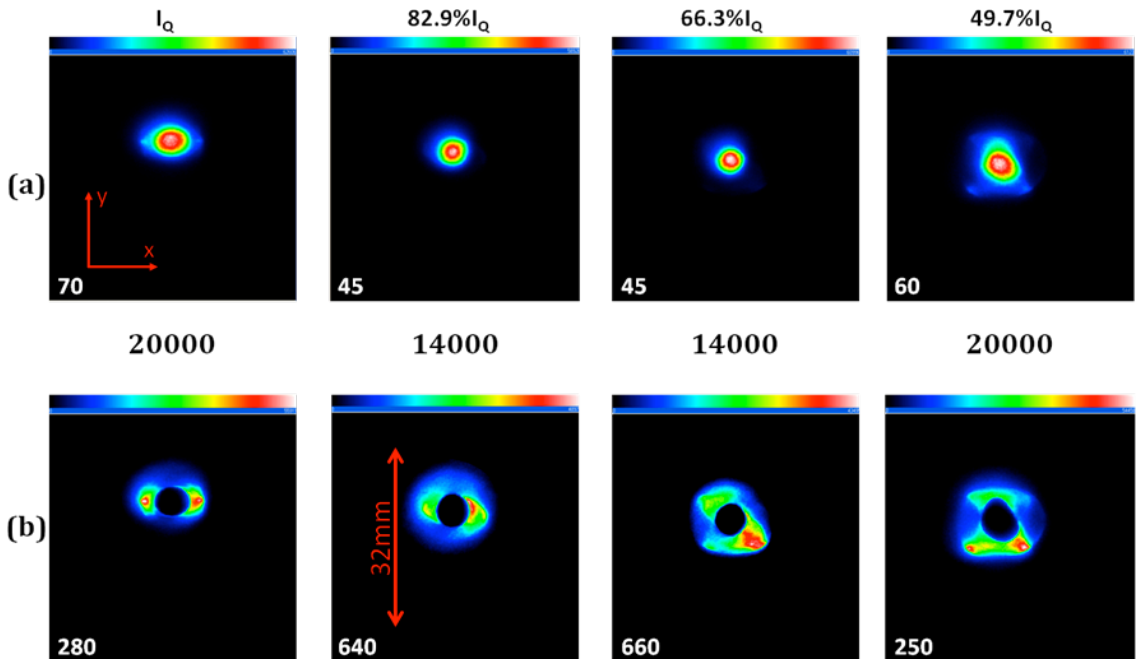

Figure 3.14: Comparison of core and halo with changing quadrupole strength;
(a) unmasked beam picture; (b) masked beam picture.

## 4. Conclusions and Outlook

In this paper, we have presented a new high dynamic range method to image beam halo using a DMD. We have designed a generic two channel optical system that provides control of the focus and magnification in each channel and incorporates two techniques to compensate for the special properties of the DMD.

We have demonstrated that 1) the DMD has excellent extinction; 2) the quality of the beam image reflected from the DMD is unaffected by the diffraction effects produced by the DMD itself; and 3) the resolution of the DMD optics is equivalent to that of comparable system in which a mirror replaces the DMD.

In addition we have measured the point spread function of our optical system and demonstrated a dynamic range of ~$10^5$ with this method, using a tightly focused UMER electron beam imaged with a simple phosphor screen. This DR matches that achieved in earlier experimental results obtained using the DMD with a laser only.

Moreover, we have used the DMD to create adaptive optical masks to block out the beam core, thus allowing us to observe the halo of the electron beam at UMER. We have shown the flexibility of this method to mask beam cores with different shapes and observed halo formation produced by varying the bias voltage in the electron gun and the strength of an upstream quadrupole magnet.

Our measurements demonstrate the quality and usefulness of the DMD masking method for high dynamic range beam imaging. This technique can readily be applied to any accelerator or light source and provides a new tool for the study of halos and beam dynamics. Using a smaller, higher intensity beam, it should be possible to achieve an even larger DR than demonstrated by our measurements at UMER. And in fact, preliminary measurements at the JLAB FEL accelerator using a DMD to image the beam non-invasively with optical synchrotron radiation, have demonstrated a dynamic range exceeding $10^5$.

Further halo studies at UMER and other accelerator facilities using upgrades to our current imaging system are already planned and will further improve the dynamic range and quality of the beam images obtained with this unique monitor. The results from these studies, when completed, will be presented in future works.